\documentclass[cameraready]{cseminar}

\usepackage[breaklinks]{hyperref}
\usepackage[hyphenbreaks]{breakurl}

\begin{document}


\title{A Survey of Augmented Reality Navigation}

\author{Gaurav Bhorkar\\
	\texttt{Gaurav.Bhorkar@aalto.fi}}
\maketitle


\begin{abstract}
  Navigation has been a popular area of research in both academia and industry. Combined with maps, and different localization technologies, navigation systems have become robust and more usable. By combining navigation with augmented reality, it can be improved further to become realistic and user friendly. This paper surveys existing researches carried out in this area, describes existing techniques for building augmented reality navigation systems, and the problems faced. 

\vspace{3mm}

\end{abstract}


\section{Introduction}

For centuries, humans have continuously tried to find ways to improve navigation. Historically using maps and compasses during the age of exploration or, in current times, the use of Google maps to find the quickest way to reach their destination. Navigation systems play a major role in our day-to-day lives. Today's systems use satellite navigation technology for positioning and display a path to users' destination. They usually provide turn-by-turn navigation instructions and occasionally a 3D rendering of the path. However, they run on a specialized device (usually a smartphone or a small computer) and require the user to divert their attention in order to interact with the software. This is often is the cause of user dissatisfaction and sometimes accidents. 

Augmented reality (AR) is a technology that "immerses" virtual objects into the real world. AR tries to mix real world and the virtual world in such a way that both real and virtual objects appear to the user to be in the same space. R. Azuma \cite{azuma1997survey} defined AR as systems that - "combine real and virtual information, interact at real time, and are registered in three dimensions". A good example of an AR system is a fighter aeroplane cockpit, where information such as target-lock, navigation, and aeroplane parameters, is projected on the cockpit glass such that it "augments" the view outside. Since information is augmented with the real world, the user's focus is not diverted to a separate device in order to seek information.

An augmented reality navigation system is a system that uses augmented reality technology in a navigation system. Essentially, the system combines navigation information which is virtual with the real world objects. Such systems can be applied to a variety of scenarios such as vehicular navigation, pedestrian navigation, indoor navigation, etc. In all of these scenarios, the aim is to help user reach the destination more efficiently by augmenting the real world with navigation information. For example, showing navigation markers on the windshield of the car or augmenting the video camera output of a smartphone with the navigation path, etc.

This paper surveys the research in this area and describes the techniques for combining augmented reality with navigation, state-of-art types of systems, opportunities and the challenges faced. Section \ref{sec:CurrentNavSys} points out limitations in current navigation systems. Section \ref{sec:ARNavigation} introduces an AR navigation system and also lists technology considerations, types of systems, opportunities and challenges in respective subsections. Finally, we put concluding remarks in section \ref{sec:conclusion}


\section{Problems of Conventional Navigation Systems}
\label{sec:CurrentNavSys}
Currently, popular navigation systems that are on the market are smartphone applications or standalone systems that make use of a mapping service and satellite localization. A navigation system typically displays a map on which the navigation path is shown along with the current position of the user. Along with this, the system is reactive and prompts the user by giving instruction, either visually or via audio, to follow the intended path. Although current navigation systems solve the problem of navigation efficiently, they lack in providing a good user experience. According to Narzt et. al. \cite{narzt2006augmented}, conventional navigation systems show navigation information in an abstracted form by typically showing arrows pointing towards the intended direction or by providing a "bird's eye view" of the map and the intended path.  

Consider a car navigation system which typically has a screen on the dashboard which shows navigation information. The driver has to look at the display in order to see the information. This essentially distracts the driver's attention from the road and such distraction is a serious problem since it compromises the safety of the driver. In fact, research shows that driver distraction due to navigation systems has lead to several accidents. P. Green \cite{green2000crashes} compiled a paper listing the number of crashes in Japan from January-November, 1999 due to various navigation system tasks. The research shows that when the driver is using a navigation system, the majority of the crashes happen when the driver is "looking" at the navigation display. Moreover, the facts and statics page of the official US Government website for distracted driving states the number of people killed in USA because of distracted driving as over 3000 in 2014 \cite{distracteddriving}. The website also lists the use of a navigation system as a major distraction. This type of distraction can be prevented by using augmented reality in navigation systems thereby ensuring the driver to focus on the road. 


\section{Augmented Reality Navigation}
\label{sec:ARNavigation}

\begin{figure}[h!]
  \begin{center}
    \includegraphics[width=.4\textwidth]{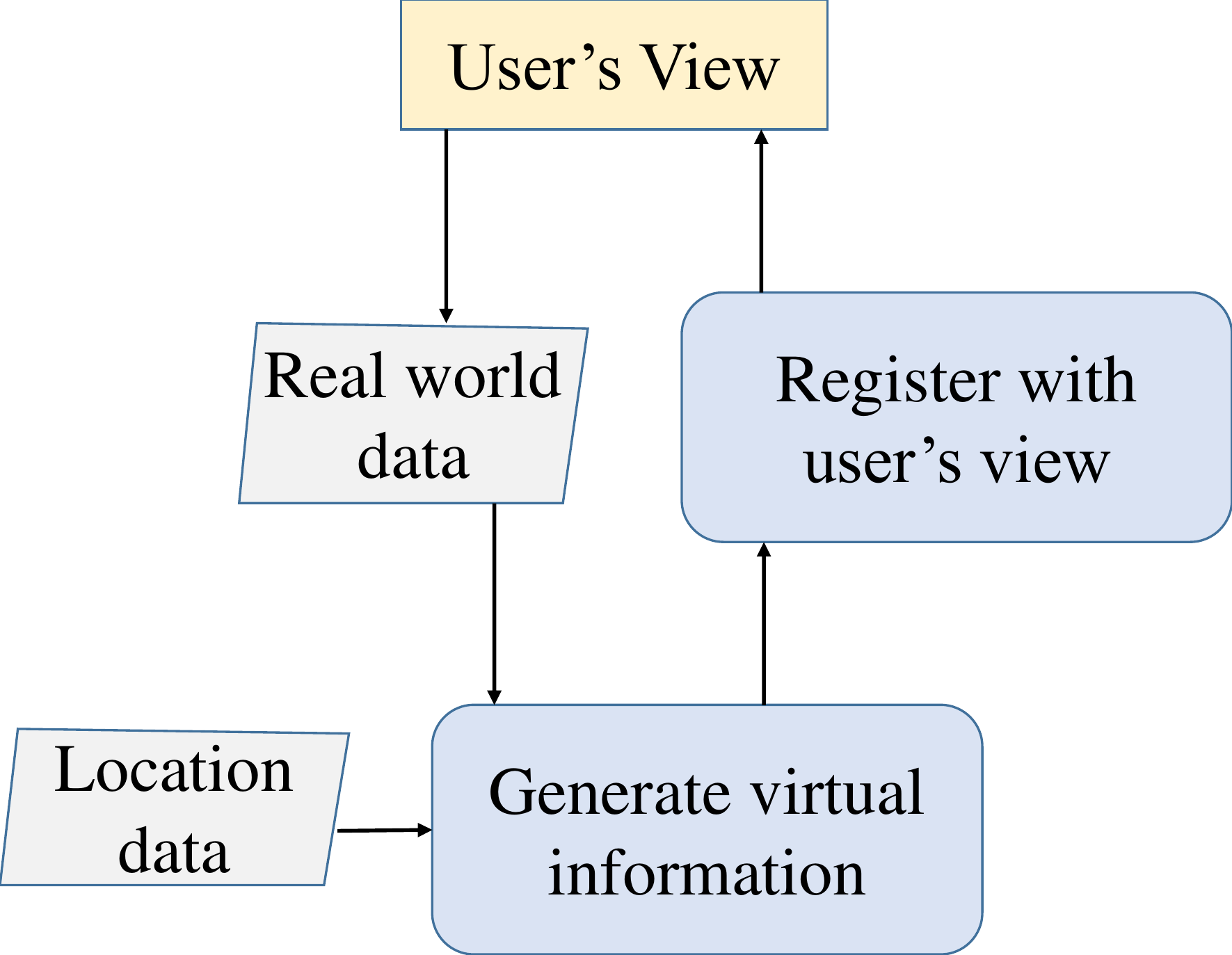}
    \caption{A simple system for AR navigation}
    \label{fig:process}
  \end{center}
\end{figure}

Generally, an augmented reality navigation would typically work in the following way (described in Figure \ref{fig:process}):
\begin{enumerate}
	\item Acquire the real world view from the user's perspective.
	\item Acquire the location information for tracking the user. This information is typically the GPS coordinates.
	\item Generate the virtual world information based on the real world view and the location information.
	\item Register the virtual information generated with the real world view and display the information to the user, thereby creating an augmented reality.
\end{enumerate}

In the following subsections we describe the technology considerations for AR navigation systems. Thereafter we introduce different types of state-of-art systems and then list the opportunities and challenges for such systems.

\subsection{Design Considerations}

The main challenges lie in combining the virtual and real worlds, and designing navigation interfaces. There are several display technology that enable augmented reality which are listed in the following section. There are also multiple navigation interfaces proposed for a good user experience.

\subsubsection{Augmented Reality Display}

How to display the augmented information to the user is a long-standing problem in AR. There are multiple display options ranging from fully virtual display to a projected display. In an AR navigation system, the display should not interfere with the user's movement. Krevelen et. al. \cite{van2010survey} describe the display technologies available for augmented reality as :

\begin{description}

	\item[Video see-through] A video see-thorough display puts a digital screen between the real world and the user. The user sees both real world and the augmented information on this screen. The real world view is captured using a camera which is then combined with augmented information and displayed on the screen, which may be in front of the user's eyes in order to offer a realistic view or it could be as a separate monitor. An example of such type of display is a head mounted display with cameras. Another example is a smartphone display with information augmented to the camera stream. 
	
	\item[Optical see-through] An optical see-through display consists of a transparent and reflective screen, also called a combiner, in front of the user's eyes. Information is projected onto this combiner by a projector. The end result is that the user is able to see virtual information and real view augmented in the same plane (the screen). Note that only the virtual information is projected on the combiner, while the real view is seen as it is since the combiner is transparent. Thus, an illusion of a combined virtual and real world is created. An example of such type of display is a fighter aeroplane cockpit glass on which virtual information is projected.
	
	\item[Projective] In a projective AR display, virtual information is directly projected on the real object instead of being shown on a digital or a see through screen. In this type of display, the users do not have to wear anything on their eyes. However, users do need to carry a projector in order to project the virtual information. This type of AR display can cover large surface areas. However, the quality of the view offered might differ depending the type of surface on which information is projected. An example of this type of a system is the SixthSense demonstrated by P. Mistry et. al. \cite{mistry2009sixthsense}. 

\end{description}

For navigation purposes, selecting an AR display may be difficult since all the display options discussed above are theoretically possible. If we consider vehicular navigation, the user's need to have free movement and minimal distraction. Narzt. et. al. \cite{narzt2006augmented} proposed a model of an optical see-through display where the windsheld of a car functions as a combiner. In the case of pedestrian and indoor navigation, a video see through display on a smartphone is a useful option.

A lot of smartphone navigation applications like HUDWAY \cite{hudway} and Sygic \cite{sygic} HUD try to reflect the smartphone screen onto the windshield. This also augments the real view with virtual information but the display is localized to a small part of the windshield. Furthermore, this technique does not process input from the driver's view and hence cannot register the virtual information with the real world to make a "true" augmented reality.

Another challenge of augmented reality is registration. Registration is the process of correctly aligning the virtual information with the real world in order to preserve the illusion of coexistence. In the case of AR in navigation, accurate registration is of utmost importance. For example, an AR navigation system pointing towards a left turn on the road, could cause confusion when there are multiple left turns because of flawed registration. Hence, even a small offset in registering virtual information is detrimental.

\subsubsection{Navigation Interface}
Navigation information can be presented to the user in many ways. A few examples include, pointing arrows towards the intended direction or by painting the road with translucent colour. 

Narzt et. al. \cite{narzt2006augmented} suggested that the interface should be designed from the driver's (user's) perspective. The path to be navigated should be painted with translucent colour. This establishes a clear path from the driver's perspective. Moreover, turns and exits that are hidden because of objects blocking the view should also be visible by overlaying the painted road on the blocking object transparently. The same points hold true for a pedestrian navigation system.

Bark et. al. \cite{bark2014personal} hold a view similar to Narzt in this regard. However, they state that the highlighted path can create confusion when it is shown passing through an object blocking the view. To remove this, they propose drawing the highlighted path in the sky so that it does not interfere with the driver's intended path. In the prototype they developed, the path is shown as a series of paper aeroplanes to guide navigation. 

A key point mentioned by Bark et. al. \cite{bark2014personal} is about augmenting the navigation interface with audio instructions as well. Thus, the system should provide navigation cues and assistance using audio. Furthermore, they developed the following guidelines for navigation interface design:
\begin{enumerate}
	\item Turns should be shown based on the driver's perspective.
	\item Upcoming instructions should be notified well in advance.
	\item The driver's view must not be cluttered and visual information must be provided only when needed.
\end{enumerate}

A global view of the path on a map is also useful along with instant augmented information. This is similar to what is shown in car racing video games such as Need for Speed. In this case, the driver has more information and can take driving decisions well in advance. Kim et. al. \cite{kim2009simulated} demonstrate a similar technique in a simulation developed for elderly drivers.
 
Along with navigation information, the user can also receive with context specific information. For example, while driving, the user could be shown the information about the nearest petrol stations, restaurants, or traffic congestion information on other routes. But it is also important to note that while context specific information is useful, it should not block the the user's view or clutter the display with excessive information.

\subsection{Different AR Navigation Systems}

This section describes the different scenarios where augmented reality can be used with respect to navigation. Traditional navigation systems are generally used for vehicular navigation and pedestrian navigation. Augmented reality can greatly enhance the user experience of these navigation services. Indoor navigation is also a good candidate for implementing AR. Other uses of AR navigation systems, for instance defence purposes (Aircraft, Naval, etc.) lie outside the scope of this paper.


\subsubsection{Car Navigation}
\label{sec:CarNav}
The main goal of using augmented reality in a car navigation system is to maintain the driver's attention on the road and solve the problems associated with traditional navigation systems as discussed in section \ref{sec:CurrentNavSys} above. Narzt et. al. \cite{narzt2006augmented} proposed a paradigm for car navigation which allows the user to be free from additional AR apparatus. This is achieved by embedding the apparatus into the user's environment, that is, the car. They propose the following paradigm for the system:

\begin{enumerate}
	\item A car windscreen to superimpose navigation information on the real visuals. 
	\item A camera to track the driver's eye position and make adjustments to the windshield display for proper overlay of the virtual objects.
	\item Tracking equipment that determines the position of the car by means of GPS and other sensors.
\end{enumerate}

\begin{figure}[h!]
  \begin{center}
    \includegraphics[width=0.5\textwidth]{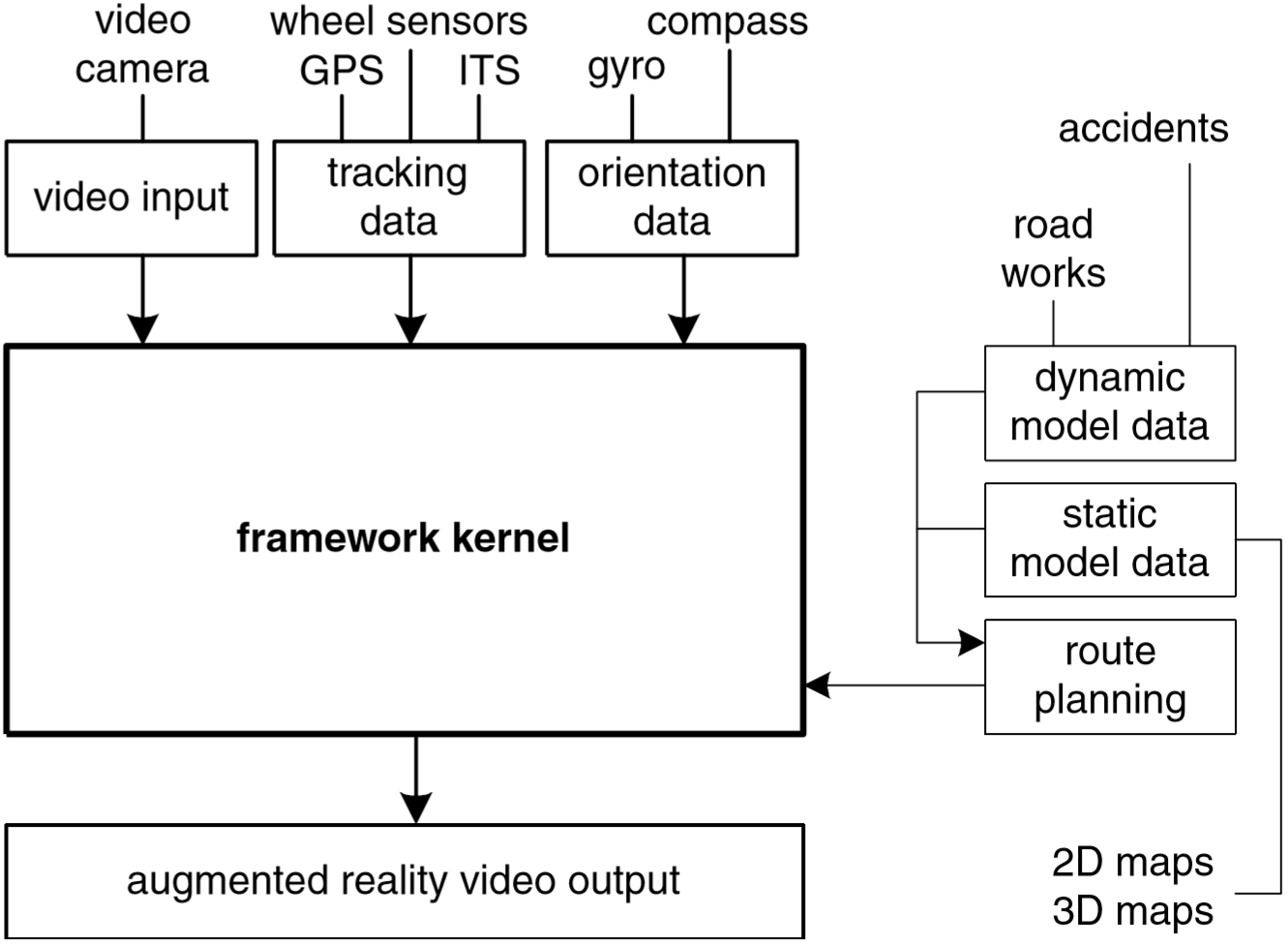}
    \caption{Narzt's AR navigation system \cite{narzt2006augmented}}
    \label{fig:narztsystem}
  \end{center}
\end{figure}

\begin{figure}[h!]
  \begin{center}
    \includegraphics[width=0.45\textwidth]{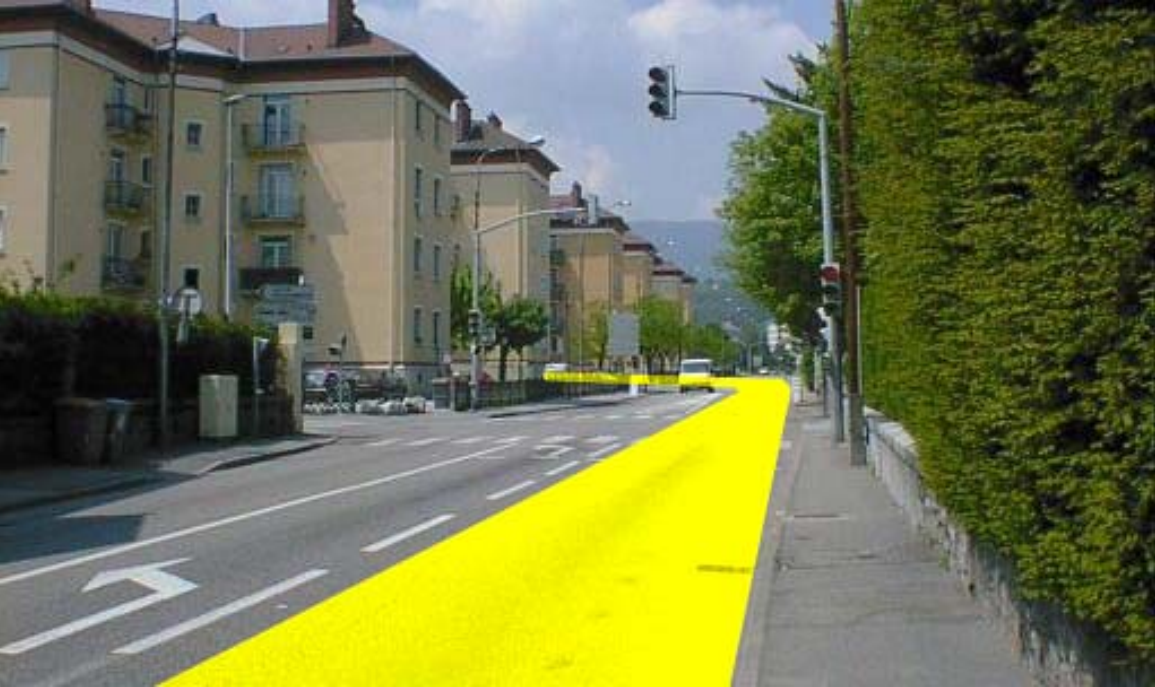}
    \caption{Sample output from Narzt's system \cite{narzt2006augmented}}
    \label{fig:narztsample}
  \end{center}
\end{figure}

Other hardware includes a video camera to obtain the view from the driver's perspective. The location data is retrieved not only from traditional GPS sensors but also by alternate tracking of the vehicle using wheel sensors if GPS is unavailable. Thus the camera stream and location data are fed to the software. The software then uses the camera input, matches it with the stored maps and dynamic road data (construction, accidents, etc.), and constructs a virtual 3D image of the road. Based on the navigation path selected by the driver, the virtual 3D image is then annotated by highlighting the road. This 3D image is then passed on to the AR renderer which registers the virtual information with the real view and displays it on the driver's windshield. An overview of this system is presented in Figure \ref{fig:narztsystem}. A sample output from the experiment is shown in Figure \ref{fig:narztsample}. The technology at the time Narzt conducted his research did not allow use of the whole windshield for the display. Hence, they displayed the AR output on a video monitor placed on the dashboard. 

The system proposed by Narzt et. al. involves lot of hardware and complex software, which makes it expensive. An alternative AR navigation system is proposed by Palinko et. al. \cite{palinko2013towards} that focuses on building a low-cost system in which a matrix of LED lights is placed on the windshield which provides the navigation information. Note that, here as well, the focus is to minimize driver distraction and keep his attention on the road. In the experiment, Palinko et. al. mounted an 8 by 7 LED matrix on the windshield and used a driving simulator for testing. The LED matrix shows turns and navigation information based on information gathered in similar way proposed by Narzt et. al. The experiment gathered data about the driver's attention time on the road and compared it with traditional navigation system. The results pointed in favour of using the LED display because the driver paid more attention to the road.

The technology today is still expensive and it is difficult to use the whole windshield as an AR display. Nevertheless, various automotive and car companies such as BMW and Toyota are focusing on the development of a full augmented reality display. A recent example of a promising AR navigation technology is BMW Mini Augmented Vision, a concept demonstrated by BMW \cite{bmwglasses}. The system has a pair of glasses worn by the driver which shows navigation information on the glasses. A number of mobile applications exist on the internet which reflect the smartphone screen to provide a head up display but are mere traditional navigation applications and not functional augmented reality navigation systems.


\subsubsection{Pedestrian Navigation}

A navigation system is not just a system for vehicular navigation but it should cater for pedestrians as well. There are a multitude of smartphone applications that function both as a pedestrian navigation system and a vehicular navigation system. These systems usually overlay navigation information on the camera output. Augmented reality navigation for pedestrians typically requires extra apparatus to be carried by the user, for example a smartphone, or a complex head mounted display \cite{azuma1997survey}. The basic principle of AR navigation for pedestrians are similar to vehicular navigation systems.

The system proposed by Narzt et. al. \cite{narzt2006augmented} uses a smartphone as an AR apparatus since smartphones are ubiquitous nowadays. The same system as described in Section \ref{sec:CarNav} was ported to a smartphone. The software uses the orientation sensors to display the AR information properly as a smartphone can be oriented in any direction. The location information is gathered via GPS sensors. If GPS is temporarily unavailable, the location can be obtained by dead reckoning using the orientation and accelerometer sensor data of the smartphone. The real world view is obtained from the camera present on the smartphone. Thereafter, the generated navigation information is combined with the real world camera stream and displayed on the smartphone screen. Intensive operations such as rendering graphics, route calculations, etc. were moved to a different server. Nowadays, these operations could be moved to a separate cloud computer and the smartphone can function as a thin client.  

\begin{figure}[h!]
  \begin{center}
    \includegraphics[width=0.4\textwidth]{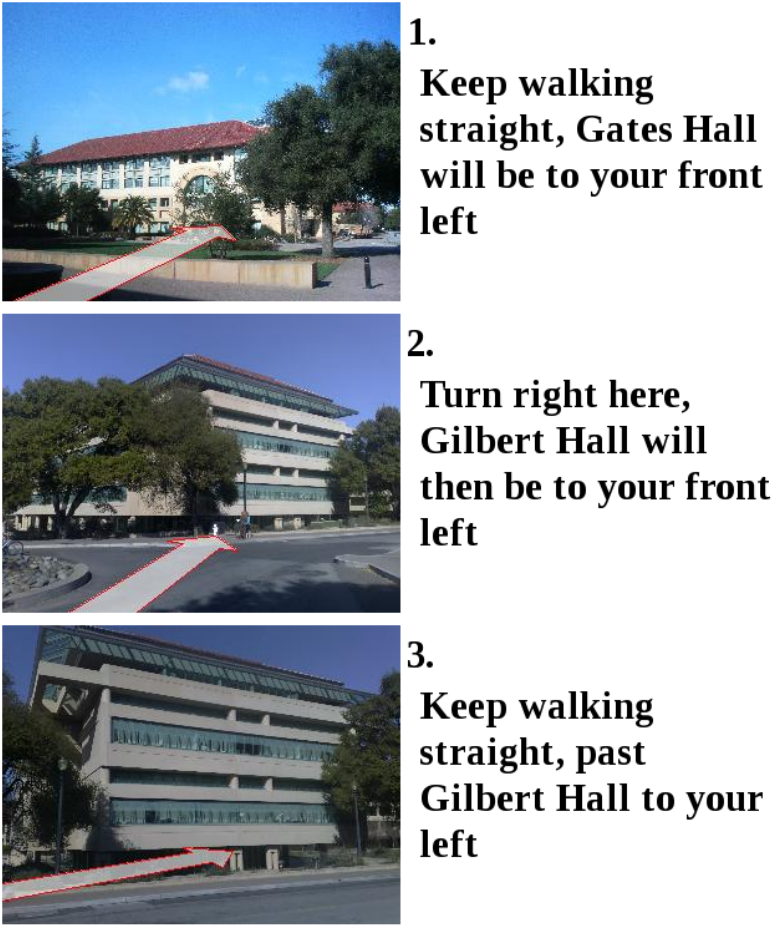}
    \caption{Sample activity from system described by Hile et. al. \cite{hile2008landmark}}
    \label{fig:hilesample}
  \end{center}
\end{figure}

An interesting example of a pedestrian navigation system is proposed by Hile et. al. \cite{hile2008landmark}. In this system, the authors have used collection of geo-tagged photographs of landmarks to create navigation instructions for a pedestrian. Once the navigation path is selected by the user, the system gathers images of landmarks along the path. Thereafter, these images are augmented with direction information using appropriately pointed arrows. Finally, the user is presented the information as a sequence of images. A sample output from this system is shown in Figure \ref{fig:hilesample}. The images used by Hile et. al. were from a privately built database of geo-tagged photos. To achieve better results, we believe that the photos can be crowd-sourced from public photo sharing websites such as Flickr, Google Picasa or even Facebook.

\subsubsection{Indoor Navigation}

Augmented reality can also be applied to indoor navigation as well. It differs from other navigation applications in that it retrieves the location of the user. It is well known that building structures attenuate GPS signals. Thus lacking GPS provided locations, indoor localization makes use of other techniques like Wi-Fi, Bluetooth, or magnetic fingerprinting. Moreover, the indoor paths have to be first mapped and stored into a database and then used as a reference for fixing the location of a user. There are a few promising technologies in indoor localization: two prominent ones are iBeacon \cite{ibeacon}, from Apple and IndoorAtlas \cite{indooratlas} from the company of same name. Some technologies make use of physical references such as QR codes for getting non-continuous location. Once the location is retrieved, it can then be used along with the camera input to create an augmented reality indoor navigation system. 

\begin{figure}[h!]
  \begin{center}
    \includegraphics[width=0.48\textwidth]{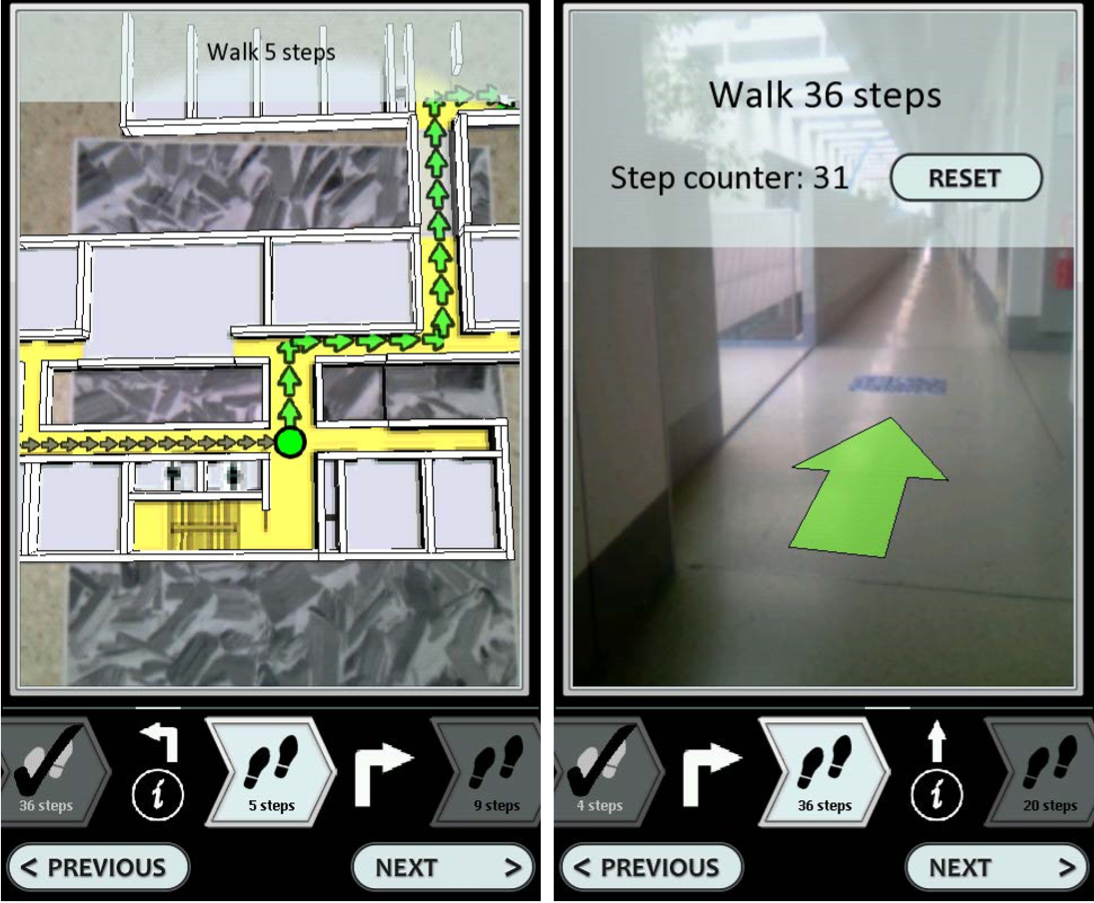}
    \caption{Activity generated by the system described by Mulloni et. al. \cite{mulloni2011handheld}}
    \label{fig:mullonisample}
  \end{center}
\end{figure}

A. Mulloni et. al. \cite{mulloni2011handheld} propose a hand-held activity-based AR indoor navigation system. The system does not track the user's location continuously, instead it uses information points (way-points) printed on the floor to detect the location of user. Once the user selects the navigation path, the system builds an activity specifying the number of steps and the turns to take. The navigation information is registered in 3D on the smartphone by superimposing the camera stream. For example, the user first selects a starting way-point by scanning the QR code on the floor and the destination as well. Thereafter, the system generates activities as shown in Figure \ref{fig:mullonisample}. The real time location of the user is calculated by dead reckoning, that is, by calculating the number of steps taken by the user from the last way-point. Since dead reckoning suffers from reduced accuracy over time, the user needs to update his/her location by scanning another way-point on the way.

\begin{figure}[h!]
  \begin{center}
    \includegraphics[width=0.5\textwidth]{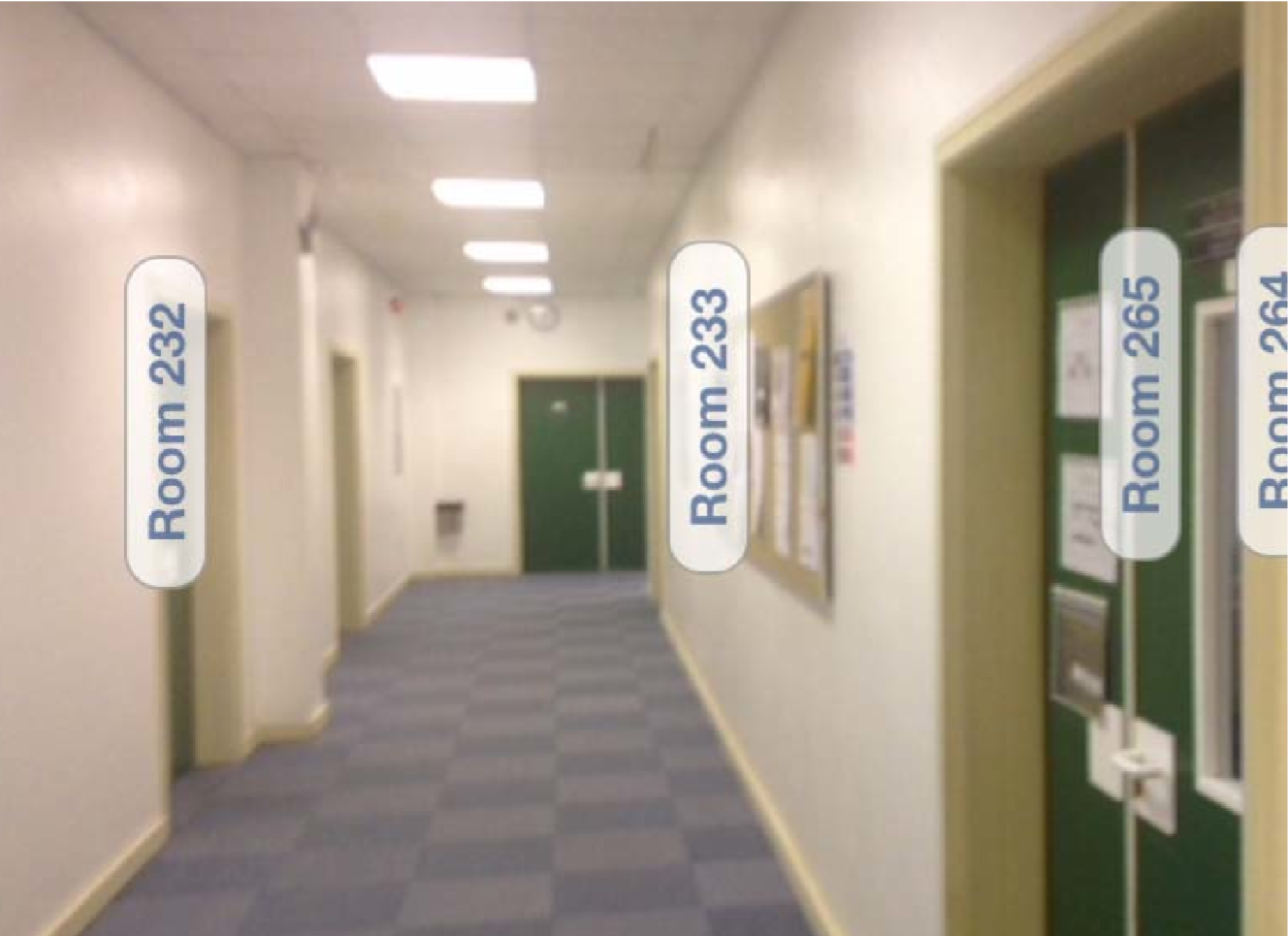}
    \caption{"360-degree points-of-interest" on system described by B. Al Delail et. al. \cite{al2013indoor}}
    \label{fig:aldelailsample}
  \end{center}
\end{figure}

B. Al Delail et. al. \cite{al2013indoor} have also build a system similar to the one described by Mulloni et. al. \cite{mulloni2011handheld}. Instead of an activity based navigation scheme, the system shows the navigation instructions in real time. The real time location is similarly retrieved using dead reckoning based on measuring the footsteps. An interesting point in the system developed is "360-degree points-of-interest" view. The system shows points of interest around the user when the camera is moved as shown in Figure \ref{fig:aldelailsample}. The points of interest are retrieved based on the user's location and are registered on the camera stream using data from the compass, gyroscope, and the accelerometer sensors. 

\subsection{Opportunities and Challenges}

Gabbard et. al. \cite{gabbard2014behind} have listed a few opportunities for augmented reality navigation systems apart from the traditional way-finding which includes driver safety where the driver is notified about information such as accident prone spots, collision warnings, warnings about changing lanes, speed limits, zebra crossings, etc. The paper also mentions showing social media information such as nearby drivers, and context specific geographical data, or important places. 

Palinko et. al. \cite{palinko2013towards} surveyed drivers and concluded that a majority of the users found the augmented reality system, despite the needed improvements, is better than the current lot of navigation devices. Similarly, a simulated system developed by Kim et. al. \cite{kim2009simulated} surveyed elderly drivers for usability. The authors conclude that the AR system significantly reduced errors in navigation and issues related to divided attention. Augmented reality applications for vehicular, pedestrian or indoor navigation are very interactive and also realistic. Moreover, combining navigation information with context specific information immensely improves the user experience. For example, consider moving through a mall using an AR navigation application on your smartphone which shows all the discounts available on the shop you pass by. 

There are some disadvantages of using an AR navigation system as well. Narzt et. al. \cite{narzt2006augmented} mentions that the user experience depends on the degree of virtual information that is augmented with the real world. The navigation information should not block the real world view of the user. This may also lead to safety issues when the AR view of the user is too intrusive to allow him/her to focus on the road. Hence, interface design of an AR navigation display is very important. AR navigation could lead to privacy concerns as well. Since the systems take real world input via a camera, other users would find it uncomfortable being surrounded by many cameras.


\section{Conclusion}
\label{sec:conclusion}

This paper presented a solid literature review of the AR navigation techniques including design considerations and different types of system. It also presents various opportunities and challenges for using AR in navigation. 

Augmented reality is a very powerful mechanism to improve user experience in navigation applications. Apart from the user experiences, it contributes immensely to the safety of driving. Surveys of the research literature suggest that AR greatly increases driver's attention. Although current technologies are still far from a full AR experience, research is moving fast and several institutes and companies like BMW, Pioneer, and Toyota are working on development in AR navigation. It is important to note that the use of AR in navigation is largely dependent on advances in the field of augmented reality. Also, the acceptance of AR navigation will depend on a better user experience. It will be interesting to see how this technology shapes in the future.



\bibliography{seminar-paper}
\end{document}